\documentstyle[12pt]{article}

\def\0{\over } \def\1{\vec }     \def\2{{1\over2}} \def\4{{1\over4}}
\def\5{\bar }  \def\6{\partial } \def\7#1{{#1}\llap{/}}
\def\8#1{{\textstyle{#1}}}       \def\9#1{{\bf {#1}}}
 \def\llp{\hbox to 0pt{\hss /\hskip1.5pt}}
\def\llo{\hbox to 0.2pt{\hss /}} \def\llq{\hbox to 0pt{\hss /\hskip0.5pt}}
\def\so{\supset\hbox to 0pt{\hss $\displaystyle -$\hskip1pt}}

\def\<{\langle } \def\>{\rangle }

\let\nn=\nonumber
\def\bea{\begin{eqnarray}} \def\eea{\end{eqnarray}}
\def\beann{\begin{eqnarray*}} \def\eeann{\end{eqnarray*}}
\def\beq{\begin{equation}} \def\eeq{\end{equation}}

\date{}
\title{
{\large\rm DESY 94-107}\hfill{\large\tt ISSN 0418-9833}\\
{\large\rm June 1994}\hfill\vspace*{3.5cm}\\
Probing Lumps of Wee Partons\\
in Deep Inelastic Scattering}
\author{W. Buchm\"{u}ller\\
{\normalsize\it Deutsches Elektronen-Synchrotron DESY, 22603 Hamburg, Germany}
\vspace*{3.5cm}\\
}
\begin{document}

\setlength{\baselineskip}{18pt}
\maketitle
\begin{abstract}
\noindent
Recently, the ZEUS collaboration has reported on several remarkable
properties of events with a large rapidity gap in
deep inelastic scattering. We suggest that the mechanism
underlying these events
is the scattering of electrons off lumps of wee partons inside the proton.
Based on an effective lagrangian approach the $Q^2$-, $x$- and
$W$-distributions are evaluated. For sufficiently small invariant mass of
the detected hadronic system, the mechanism implies leading
twist behaviour.
The $x$- and $W$-distributions are determined by the Lipatov
exponent which governs the behaviour of parton densities at small $x$.
\end{abstract}
\newpage
In a recent publication \cite{zeus1}, the ZEUS collaboration has described
in detail properties of events with a large rapidity gap in
deep inelastic scattering (RGDIS). The events, whose observation
was reported already some time ago \cite{zeus2,h1}, are very different
from ordinary deep inelastic scattering (DIS) where the distribution
of hadrons in the final state corresponds to a colour flow
between the scattered quark and the proton remnant. In the new class of
events the observed large rapidity gap around the proton direction
suggests that the proton remains essentially unaffected by the
scattering process and escapes undetected close to the forward direction.
The detected hadronic final state would then be the remnant of a
colour neutral part of the proton, carrying only a small momentum fraction.

A particularly striking feature of the new class of events is their
$Q^2$-dependence. Naively, one might expect that the rate of
diffractive processes, where the proton is not broken up, decreases with
increasing $Q^2$. The data, however, show the same $Q^2$ dependence
for RGDIS events and ordinary DIS events. Another important
feature is the occurence of jets in the RGDIS events, which has been
anticipated by several groups \cite{ingelman}-\cite{bruni}, following the work
of Ingelman and Schlein \cite{ingelman}. In the following, we shall try to
develop a simple picture of the mechanism underlying the RGDIS
events, which accounts for their main qualitative features.

The inelastic scattering process
\beq \label{process}
e(k)+p(P) \rightarrow e(k')+p(P')+X(P_X)\ ,
\eeq
where $P'=(1-\xi)P$, $q=k-k'$, $P_X=q+\xi P$ is characterized by four
kinematic variables: the total center-of-mass energy $\sqrt{s}$,
the squared four momentum transfer $Q^2$ and Bjorken's
scaling variable $x$, occuring in ordinary deep inelastic scattering,
\beq
s=(k+P)^2\ ,\ Q^2=-q^2=-(k-k')^2\ ,\  x={Q^2\over 2 q\cdot P}\ ,
\eeq
and the invariant mass of the detected hadronic state X,
\beq
M^2 = (q+\xi P)^2\ .
\eeq
The invariant mass of the total hadronic final state, including the
undetected proton, is
\beq \label{invmass}
W^2 = (q+P)^2 \simeq {Q^2(1-x)\over x}\ .
\eeq
In eq.\ (\ref{process}) we have neglected
the squared momentum transfer $t$ between initial and final proton.
In general, one has
\beq
\xi ={M^2 + Q^2 \over W^2 + Q^2} = x {M^2 + Q^2 \over Q^2}\ ,
\qquad  M^2 \leq \hat{s}-Q^2\ ,
\eeq
where  $\hat{s}= \xi s$.
The ZEUS data cover the kinematic range $Q^2 > 10$ GeV$^2$,
$W^2 \gg Q^2, M^2$
and, hence, $x < \xi \ll 1$.

The observed ratio of RGDIS events over ordinary
DIS events is $\cal{O}(10\%)$. We conclude that an electron, which
probes the interior of the proton, frequently separates a
colour neutral part with small momentum fraction, without breaking
the proton into pieces. Visualizing a fast moving proton at $Q^2 > 10$ GeV$^2$,
or $\Delta X_{\perp} < 0.06$ fm, as a bunch of valence quarks, sea quarks and
gluons,
this means that a lump of ``wee'' partons \cite{feynman}, i.e.,
soft gluons and sea quarks, is rather easily stripped
off. In the scattering process, this colour neutral hadronic state
of size $1/Q$ is turned into a new
hadronic state of similar size but with invariant mass $M > 0$, which then
fragments. The virtual photon, which can fluctuate into hadrons, acts
like a disk of radius $1/Q$ \cite{bjorken},
which presses  a droplet of wee partons out of the proton.

Let us try to make this picture quantitative
by means of an effective lagrangian,
where the initial and final hadronic systems are represented by fields
$\sigma$ and $V$, respectively. The simplest hypothesis is to
describe the wee parton cluster by a massless scalar field $\sigma$,
carrying vacuum quantum numbers, and to choose for $V$ a massive vector field.
The interaction with the photon should then be dominated by the gauge invariant
operators of lowest dimension. One easily verifies that,
up to terms vanishing by the equations
of motion, there is a unique dimension five operator,
\beq \label{effint}
\cal{L}_I = - {1\over 4\Lambda} \sigma(x) F_{\mu\nu}(x) V^{\mu\nu}(x)\ ,
\eeq
where $F_{\mu\nu}=\partial_{\mu}A_{\nu}-\partial_{\nu}A_{\mu}$
and $V_{\mu\nu}=\partial_{\mu}V_{\nu}-\partial_{\nu}V_{\mu}$
are the field strength tensors of the photon field
and the hadronic vector field, respectively. Note, that
the choice of a scalar field for the hadronic final state would require
an interaction operator of dimension six, since the masses of initial and
final state are different. According to the qualitative picture of the
scattering process described above, the length
$1/\Lambda$ will be identified
with the transverse size of the photon, i.e., $1/\Lambda \sim 1/Q$.

The differential cross section for the production of a
hadronic vector state with fixed mass $M$ is now easily evaluated. The
interaction term (\ref{effint}) yields the $\gamma \sigma V$-vertex
$-i(g_{\mu\nu}q\cdot P_X - q_{\nu} P_{X\mu})/2\Lambda$,
where $q$ and $P_{X}$
are the momenta of the photon ($A_{\mu}$)and the hadronic final state
($V_{\nu}$) , respectively.
 From the matrix element
for the quasi-elastic process $e + \sigma \rightarrow e' + V(M)$
one obtains the differential cross section
\beq \label{xparton}
{d\tilde{\sigma} \over d\xi dQ^2} = {\pi \alpha^2\kappa^2 \over 4 Q^4}
\left(1 - y + {y^2\over 2} - 2{Q^2 M^2\over (Q^2+M^2)^2} y^2
\right) f_{\sigma}(\xi,Q^2)\ ,\
y = {q\cdot P\over k\cdot P}\ .
\eeq
Here $f_{\sigma}(\xi,Q^2)$ is the density of wee parton clusters with
momentum fraction $\xi$, and
we have defined $1/\Lambda = e\kappa/Q$, as discussed above.
$\kappa$ is an unknown constant and $e$ is the electromagnetic coupling.
Note, that the derivatives in the interaction (\ref{effint}) generate
one power of $Q^2$, which cancels the inverse power of $Q^2$
stemming from $1/\Lambda \sim 1/Q$. In order to obtain the complete
differential cross section for RGDIS events eq.\ (\ref{xparton}) has
to be multiplied with the density $\rho(M^2)$ of massive vector states.

What is the probability density $f_{\sigma}(\xi, Q^2)$ of finding a
colour neutral lump of wee partons with momentum fraction $\xi$
inside the proton? Clearly, at least two gluons or a quark and an antiquark
with momentum fractions $\cal{O}(\xi)$ are needed. Assuming that the two-parton
component gives the dominant contribution, one obtains
\beq \label{lump}
f_{\sigma}(\xi,Q^2) = \int_{r\xi}^{(1-r)\xi}d\xi'\left(f_g
g(\xi',Q^2)g(\xi-\xi',Q^2) + f_S S(\xi',Q^2) S(\xi-\xi',Q^2)\right)\ ,
\eeq
where $f_g$, $f_S$ and the fraction $r<1$ are constants.
At values of $\xi$ below $\sim 10^{-2}$, one has \cite{mrs}
\beq \label{martin}
g(\xi,Q^2) \simeq A_g \xi^{-1-\lambda}\ ,\
S(\xi,Q^2) \simeq A_S \xi^{-1-\lambda}\ ,\ A_S \ll A_g\ .
\eeq
Here $\lambda = \lambda(Q^2) > 0$ is the Lipatov exponent \cite{lipatov}.
Eqs.\ (\ref{lump}) and (\ref{martin}) imply
\beq \label{lump1}
f_{\sigma}(\xi,Q^2) = C \xi^{-1-2\lambda}\ ,
\eeq
where the constant $C$ is a function of $f_g$, $f_S$, $r$, $A_g$ and $A_S$.
Note, that for
$\lambda > 0$ the density of wee parton clusters is more
singular than the desities of the individual partons.
The hypothesis of a distribution of colourless lumps of wee partons
inside the proton is reminiscent of ``preconfinement'', the formation
of colourless clusters of quarks and gluons in the evolution of jets,
as suggested by Amati and Veneziano \cite{amati}.

A plausible ansatz for the spectrum of massive vector states is provided by
generalized vector meson dominance (GVD)\footnote{For a review,
see ref.\ \cite{yennie}.}. As an example, we use the model
of Sakurai and Schildknecht \cite{sakurai} for the spectral densities
of vector states coupling to the photon, which is compatible with
scaling in deep inelastic electron-proton scattering. The densities of
transverse and longitudinal vector states are given by \cite{sakurai}
\bea \label{spectrum}
\rho_T(M^2)&=&\sum_{V=\rho,\omega,\phi}r_V \delta(M^2-m_V^2)+r_c
{M_0^2\over M^4} \theta(M^2-M_0^2)\ ,\\
\rho_L(M^2)&=& {1\over 4}\sum_{V=\rho,\omega,\phi}r_V \delta(M^2-m_V^2)\ ,\nn
\eea
where $r_c \simeq 0.22$ and $M_0\simeq 1.4$ GeV.

The differential cross section (\ref{xparton}) can be decomposed into
transverse and longidudinal components in the standard manner,
\bea
\tilde{\sigma}_L(x,Q^2) &=& {4\pi^2 \alpha\over Q^2(1-x)} \tilde{F}_L
={4\pi^2 \alpha\over Q^2(1-x)}{\kappa^2\over 4}{Q^2 M^2\over (Q^2+M^2)^2}
\xi f_{\sigma}(\xi,Q^2)\ ,\\
\tilde{\sigma}_T(x,Q^2) &=& {4\pi^2 \alpha\over Q^2(1-x)}
(\tilde{F}_2 - \tilde{F}_L)
={4\pi^2 \alpha\over Q^2(1-x)}{\kappa^2\over 16}
{(Q^2-M^2)^2\over (Q^2+M^2)^2}\xi f_{\sigma}(\xi,Q^2)\ .\nn
\eea
Here we have used $d\xi\simeq \xi/x dx$.
The complete diffractive cross sections are now obtained by integrating
over the mass spectra,
\bea \label{xdiff}
\sigma^D_T(x,Q^2) &=& {4\pi^2 \alpha\over Q^2(1-x)}{\kappa^2\over 16}
\int dM^2 {(Q^2-M^2)^2\over (Q^2+M^2)^2}\rho_T(M^2) \xi f_{\sigma}(\xi,Q^2)\
,\\
\sigma^D_L(x,Q^2) &=& {4\pi^2 \alpha\over Q^2(1-x)}{\kappa^2\over 4}
\int dM^2 {Q^2 M^2\over (Q^2+M^2)^2 }\rho_L(M^2) \xi f_{\sigma}(\xi,Q^2)\ .\nn
\eea
The corresponding diffractive differential cross section reads
\bea
{d\sigma^D \over dx dQ^2 dM^2} &=& {\pi \alpha^2\kappa^2 \over 4 x Q^4}
\left[\left(1 - y + {y^2\over 2}\right)
\left({(Q^2-M^2)^2\over (Q^2+M^2)^2} \rho_T(M^2) +
4{Q^2 M^2\over (Q^2 + M^2)^2} \rho_L(M^2)\right)\right.\nn \\
& & \qquad \left.- 2{Q^2 M^2\over (Q^2+M^2)^2} y^2 \rho_L(M^2)
\right]\xi f_{\sigma}(\xi,Q^2)\ .
\eea
In our calculation we have neglected a possible transverse momentum of
the final state $X$ with respect to the proton direction. The dependence
on the transverse momentum is needed in order to compute
the differential cross section $d\sigma /dt$, where
$t=(P-P')^2$ is the squared momentum transfer of the proton.

The cross sections for RGDIS events are determined by
eqs.\ (\ref{lump1}), (\ref{spectrum}) and (\ref{xdiff}). In the case
$Q^2 \gg M_0^2$, where the continuum states dominate, one obtains
\bea \label{diffinel}
\sigma_T^D (x,Q^2) &=& {4\pi^2 \alpha\over Q^2(1-x)}
{\kappa^2 C r_c\over 16} x^{-2\lambda}\left(1+
\cal{O}\left({M_0^2\over Q^2}\right)\right)\ ,\\
\sigma^D_L (x,Q^2) &=& 0\ .\nn
\eea
For comparison, the cross sections for ordinary DIS events are,
neglecting QCD radiative corrections,
\bea \label{xinel}
\sigma^{\gamma}_T(x,Q^2) &=& {4\pi^2\alpha\over Q^2(1-x)}
\sum_q e_q^2 x \left(f_q(x,Q^2)+f_{\bar{q}}(x,Q^2)\right)\ ,\\
\sigma^{\gamma}_L(x,Q^2) &=& 0\ ,\nn
\eea
where $e_q$ is the electric charge of the quark $q$.
{}From eqs.\ (\ref{diffinel}) and (\ref{xinel}) one immediately reads off
that the ratio of RGDIS events over ordinary DIS events,
\beq
r(x,Q^2) = {\sigma^D_T(x,Q^2)\over \sigma^{\gamma}_T(x,Q^2)}\ ,
\eeq
is independent of $Q^2$, up to logarithmic corrections and terms
$\cal{O}(M^2_0/Q^2)$. Hence, the RGDIS events have leading twist behaviour.

The $x$-dependence of the ratio $r(x,Q^2)$ can be obtained
from eqs.\ (\ref{martin}),
(\ref{diffinel}) and (\ref{xinel}).
At small values of $x$ below $\sim 10^{-2}$, one has
$f_q(x,Q^2)\simeq f_{\bar{q}}(x,Q^2)\simeq S(x,Q^2)$. Thus one arrives at
(cf.\ (\ref{invmass}))
\beq
r(x,Q^2) \sim x^{-\lambda} \sim W^{\lambda}\ .
\eeq
Hence, the $x$- and $W$-distributions of the RGDIS events are completely
determined by the Lipatov exponent $\lambda$.

One may be tempted by
the form of the spectral densities $\rho_T(M^2)$ and $\rho_L(M^2)$ to apply
the diffractive cross sections eqs.\ (\ref{xdiff}) also to the
exclusive production of vector mesons. However, this procedure appears
doubtful, since the lump of wee partons described by the field $\sigma(x)$
consists primarily of gluons
(cf.\ (\ref{lump}),(\ref{martin})), whereas the production cross section
of vector mesons is determined
by the quark-antiquark component of the wavefunction. A naive application of
eqs.\ (\ref{spectrum}) and (\ref{xdiff}) to $\rho$-production would lead to
$\sigma_T^{\rho} \sim F^{\rho}_T/Q^2$ and
$\sigma_L^{\rho} \sim F^{\rho}_L m^2_{\rho}/Q^4$,
where the form factors $F^{\rho}_{T,L}(Q^2/m_{\rho}^2)$
account for the transition from
the final hadronic vector state of size $1/Q$ to a physical $\rho$-meson.
Assuming the same form factor for transverse and longitudinal polarization,
one would obtain $\sigma^{\rho}_L \sim \sigma^{\rho}_T/Q^2$,
in contradiction to
predictions based on non-perturbative \cite{donlan} and perturbative
\cite{rys, brod} two-gluon exchange.

In summary, we have developed a simple model which can account for the
qualitative features of the RGDIS events.
Starting point is the hypothesis that the wee partons inside the proton
form colourless clusters, which are represented by a scalar field carrying
vacuum quantum numbers.
The observed leading twist
behaviour then essentially follows from dimensional analysis and
gauge invariance. Also important is the dominance of hadronic final
states with masses small compared to $Q$. Hence, the leading twist
behaviour should disappear as $Q^2$ approaches the continuum threshold
$M_0^2$. The density of wee parton clusters has been calculated in terms
of the gluon and sea quark densities inside the proton. At small values
of $x$, where sea quarks and gluons dominate over valence quarks,
this leads to a prediction for the $x$- and $W$-distributions of
RGDIS events in terms of the Lipatov exponent which governs the
small-$x$ behaviour of the single parton densities.

The proposed coupling between photon, and
initial and final hadronic state is an effective
description of multi-parton processes. Hence, the occurance of jets
in the final state is expected. Jet cross sections could be calculated
using the photon structure function and
the distributions of gluons and quarks inside the wee parton cluster,
which are given by eq.\ (\ref{lump}).

Experimental data have led us to a description of diffractive processes
in deep inelastic scattering which is based on an effective lagrangian
for the coupling between virtual photons and clusters of wee partons.
It remains to be seen whether this ansatz can also be derived
from QCD, the theory of strong interactions.

I am grateful to G. Ingelman, G. Kramer, T. F. Walsh and P. M. Zerwas
for stimulating questions, suggestions
and comments on the manuscript. I also thank R. Klanner
and G. Wolf for helpful discussions on the diffractive events
observed in the ZEUS detector.


\begin{thebibliography}{99}
\bibitem{zeus1}
ZEUS collaboration, M. Derrick et al., preprint DESY 94-063 (1994)
\bibitem{zeus2}
ZEUS collaboration, M. Derrick et al., Phys. Lett. B315 (1993) 481
\bibitem{h1}
H1 collaboration, A. De Roeck, in Proc. of the Int. Europhys. Conf. on
High Energy Physics (Marseille, 1993) eds. J. Carr and M. Perrottet,
p. 791
\bibitem{ingelman}
G. Ingelman and P. Schlein, Phys. Lett. 152B (1985) 256
\bibitem{silva}
N. Artega-Romero, P. Kessler and J. Silva, Mod. Phys. Lett. A1 (1986) 211
\bibitem{landshoff}
A. Donnachie and P. V. Landshoff, Phys. Lett. 191B (1987) 309
\bibitem{streng}
K. H. Streng, in Proc. of the HERA Workshop (Hamburg, 1988) ed. R. D. Peccei,
p. 365
\bibitem{nikolaev}
N. N. Nikolaev and B. G. Zakharov, Z. Phys. C53 (1992) 331
\bibitem{bruni}
P. Bruni and G. Ingelman, in Proc. of the Int. Europhys. Conf. on
High Energy Physics (Marseille, 1993) eds. J. Carr and M. Perrottet,
p. 595
\bibitem{feynman}
R. P. Feynman, Photon-Hadron Interactions (Benjamin, Reading, 1972)
\bibitem{bjorken}
J. D. Bjorken, in Proc. of the 1971 Int. Symp. on Electron and Photon
Interactions at High Energies (Ithaca, N. Y., 1971) ed. N. B. Mistry,
p. 13
\bibitem{mrs}
A. D. Martin, R. G. Roberts and W. J. Stirling, Phys. Rev. D47 (1993) 867
\bibitem{lipatov}
E. A. Kuraev, L. N. Lipatov and V. S. Fadin, Sov. Phys. JETP 45 (1977) 199;\\
Ya. Ya. Balitskii and L. N. Lipatov, Sov. J. Nucl. Phys. 28 (1978) 822
\bibitem{amati}
D. Amati and G. Veneziano, Phys. Lett. 83B (1979) 87
\bibitem{yennie}
T. H. Baur, R. D. Spital, D. R. Yennie and F. M. Pipkin, Rev. Mod. Phys.
50 (1978) 261
\bibitem{sakurai}
J. J. Sakurai and D. Schildknecht, Phys. Lett. 42B (1972) 216
\bibitem{donlan}
A. Donnachie and P. V. Landshoff, Phys. Lett. 185B (1987) 403;
Nucl. Phys. B311 (1988) 509
\bibitem{rys}
M. G. Ryskin, Z. Phys. C57 (1993) 89
\bibitem{brod}
S. J. Brodsky et al., preprint SLAC-PUB-6412 (1994)
\end{thebibliography}
\end{document}